# Spin-Torque-Driven Non-uniform Dynamics of an Antivortex Core in Truncated Astroid Shaped Nanomagnets

Ahmet Koral Aykin[a,b], Hasan Piskin[a,c], Bayram Kocaman[a,d], Cenk Yanik[l], Vedat Karakas[a], Sevdenur Arpaci[a,e], Aisha Gokce Ozbay[a,f], Mario Carpentieri[g], Giovanni Finocchio[h], Federica Celegato[i], Paola Tiberto[i], Sergi Lendinez[j], Valentine Novosad[j], Axel Hoffmann[j,k], Ozhan Ozatay[a,l]

[a] Boğaziçi University, Physics Department, Istanbul, Türkiye
[b] Boğaziçi University, Mechanical Engineering Department, Istanbul, Türkiye
[c] Department of Fundamental Sciences, Faculty of Engineering, Alanya Alaaddin Keykubat University, 07450, Antalya, Turkiye
[d] Siirt University, Department of Electrical and Electronics Engineering, Siirt, Turkey
[e] Northwestern University, Department of Electrical and Computer Engineering, Evanston, IL, USA
[f] TUBITAK National Metrology Institute, UME, Kocaeli, Türkiye
[g] Politecnico di Bari, Department of Electrical & Information Engineering, Bari, Italy
[h] University of Messina, Department of Mathematical and Computer Sciences, Physical Sciences and Earth Sciences, Messina, Italy
[i] Istituto Nazionale di Ricerca Metrologica (INRiM), Torino, Italy
[j] Argonne National Laboratory, Materials Science Division, Lemont, IL, USA
[k] University of Illinois Urbana-Champaign, Department of Materials Science and Engineering and Materials Research Laboratory, Urbana, IL, USA
[l] SUNUM, Sabanci University Nanotechnology Research and Application Center, 34956 Istanbul, Türkiye

## ABSTRACT

Spin textures that are not readily available in the domain structures of continuous magnetic thin films can be stabilized when patterned to micro/nano scales due to the dominant effect of dipolar magnetic interactions. Fabrication of such devices enables a thorough study of their RF dynamics excited by highly concentrated spin-polarized/pure-spin currents. For this purpose, in this study, we have employed a truncated astroid geometry to achieve stable magnetic antivortex core nucleation/annihilation which was detectable using the anisotropic magnetoresistance (AMR) at various temperatures. Furthermore, by depositing a soft magnetic thin film (20 nm thick permalloy) capped with a heavy-metal 2nm Pt layer, we were able to probe the spin orbit torque induced excitations accompanied by self-torque due to half-antivortex cores reminiscent of an isolated-antivortex, yielding GHz frequency oscillations with high quality factors (~50000). The observed RF oscillations can be attributed to a non-uniform domain wall oscillation mode close to the stable-antivortex core nucleation site as seen in micromagnetic simulations. This fundamental study of antivortex core response to spin currents is crucial for the assessment of their potential applications in high frequency spintronic devices such as reservoir computers.

Spin textures such as vortices (V) and antivortices (AV) can nucleate in soft ferromagnetic materials through the minimization of total energy as a result of the interplay between exchange energy and magnetostatic energy. While both structures have a nanometer-sized core region with magnetization pointing perpendicular to plane, they can be distinguished by their associated circular symmetry for vortices corresponding to a topological charge (vorticity or winding number) of +1 and inversion symmetry about the Bloch point for antivortices corresponding to a topological charge of –1 [1-3]. The core magnetization direction determines the polarity being 1 (up) or -1 (down), which is of potential interest for binary data storage. In addition, the direction of magnetization around the core is determined by the circulation (being a conserved quantity within a vortex but a non-conserved quantity within an antivortex) leading to magnetic charges (four alternating poles) at its edges [4-6].

While in continuous films vortices and antivortices are observed together in cross-tie domain walls [7], antivortices are usually metastable intermediate states observed in the course of the magnetization reversal process [8]. In contrast, in patterned special geometries, the shape anisotropy allows the stabilization of isolated-antivortices as well. There have been numerous reports on micromagnetic simulation results and experimental evidence for stable nucleation of isolated-antivortices in astroids [7,9], infinity-shaped structures [10,11], φ-shaped structures [12] and, pound-key-like structures [13,14] following a simple magnetic field treatment procedure. Within these possible geometries the astroid structure carries a unique advantage since the nucleated antivortex state obtained by a field treatment is determined to be quite robust to small field/current perturbations at the remanent state, allowing both reliable field and current dependent switching and radio frequency dynamics studies [15,16,17,18].

There are several ways to probe the antivortex structure including direct imaging with magnetic force microscopy (MFM) [7,12,13,17], scanning transmission X-ray microscopy [11], magneto-optical Kerr effect [14] as well as electrical detection via anisotropic magnetoresistance [4,19], all of which can be easily interpreted when combined with micromagnetic simulations. Besides, antivortex dynamic modes can be excited with several methods: 1- with small ac currents [1] which then mixes with the gyrotropic resonance signal leading to a measurable rectified voltage [20], 2- with radio frequency fields leading to spin waves detectable using micro-focused Brillouin Light Scattering [21], 3- with a static magnetic field to displace the core from equilibrium position and detect gyrotropic eigenmodes using absorption spectroscopy [5] and 4- with an in-plane rotating magnetic field to image switching dynamics using time resolved scanning transmission x-ray microscopy [11]. Moreover, they can be excited with alternative methods such as: application of a field pulse [6,9,22], suprathreshold spin polarized dc current [23], spin polarized ac current [18] and spin polarized current pulse [24]. On the detection side, spin torque ferromagnetic resonance (ST-FMR) [25] and direct measurements of the magnetoresistance oscillations with a spectrum analyzer [16] are effectively used to map out the spin torque excitations.

While in cross-tie domain walls both persistent oscillations and chaotic motion are theoretically expected [26], analytical models and micromagnetic simulations for isolated-antivortices predict three scenarios: a gyrotropic motion of the AV-core on an elliptical orbit [2,9,22], core reversal [9,23] and transition to a vortex state [24]. On the application side it is important to note that in device arrays vortex-antivortex oscillators can be synchronized via phase locking to improve the power output of the RF oscillations [15].

Spin transfer torques (STT) and spin-orbit torques (SOT) are efficient means of exciting the magnetization dynamics and reversal since they target a localized region allowing the desired high density device integration [27] with much less current requirement as compared to Oersted field driven excitations. The implementation of SOT devices is relatively facile since the spin Hall effect in a heavy-metal layer results in a transverse spin current in a simple bilayer stack. SOT can excite RF dynamics if the resulting spin torque from the transverse spin current can overcome the magnetic damping in the ferromagnetic layer to generate persistent oscillations. This was recently utilized to detect low frequency oscillations of the vortex state in a permalloy/platinum disk [25]. However, in the case of antivortices in an astroid, an additional complication arises in the interpretation of the detected oscillations owing to the possible interactions with the current that branches out to top and bottom set of rings. These interactions result in oscillations of the AV-core coupled to a closure domain wall and give rise to a non-uniform excitation of magnetization. In this sense, antivortices in an asteroid is of great potential towards efficient manipulation of magnetization for reservoir computing [28].

We have designed a truncated astroid that allows the magnetic flux closure and current flow constriction to the central region simultaneously (details of sample fabrication can be found in supplementary material). To nucleate a stable-antivortex core configuration in the remanent state at the central region of the astroid device, the samples were conditioned using in-plane AC demagnetization with 1 T max field followed by saturation in the out-of-plane direction with 1 T field (see supplementary material for further details). Magnetic force microscopy (MFM) imaging of the antivortex was performed in the lift mode with lift height of 60 nm employing a homemade magnetic tip specially designed for in-plane magnetization gradient measurements (Fig. 1-b). On the other hand, micromagnetic OOMMF simulations (see supplementary material for details) [29] suggest that the presence of interfacial DMI strength of ~1 mJ/m$^2$ for 2 nm Pt [30] and the small in-plane uniaxial anisotropy (anisotropy constant 35 J/m$^3$) facilitate the nucleation and stabilization of the antivortex state with the AV-core residing at the boundaries (Fig. 1-c black circles) which matches the experimentally observed core locations (Fig. 1-b yellow circles). It is worth mentioning that the atomic force microscopy (AFM) signal due to the topography may interfere with the magnetic signal in MFM as can be clearly seen in the gaps of the truncated astroid.

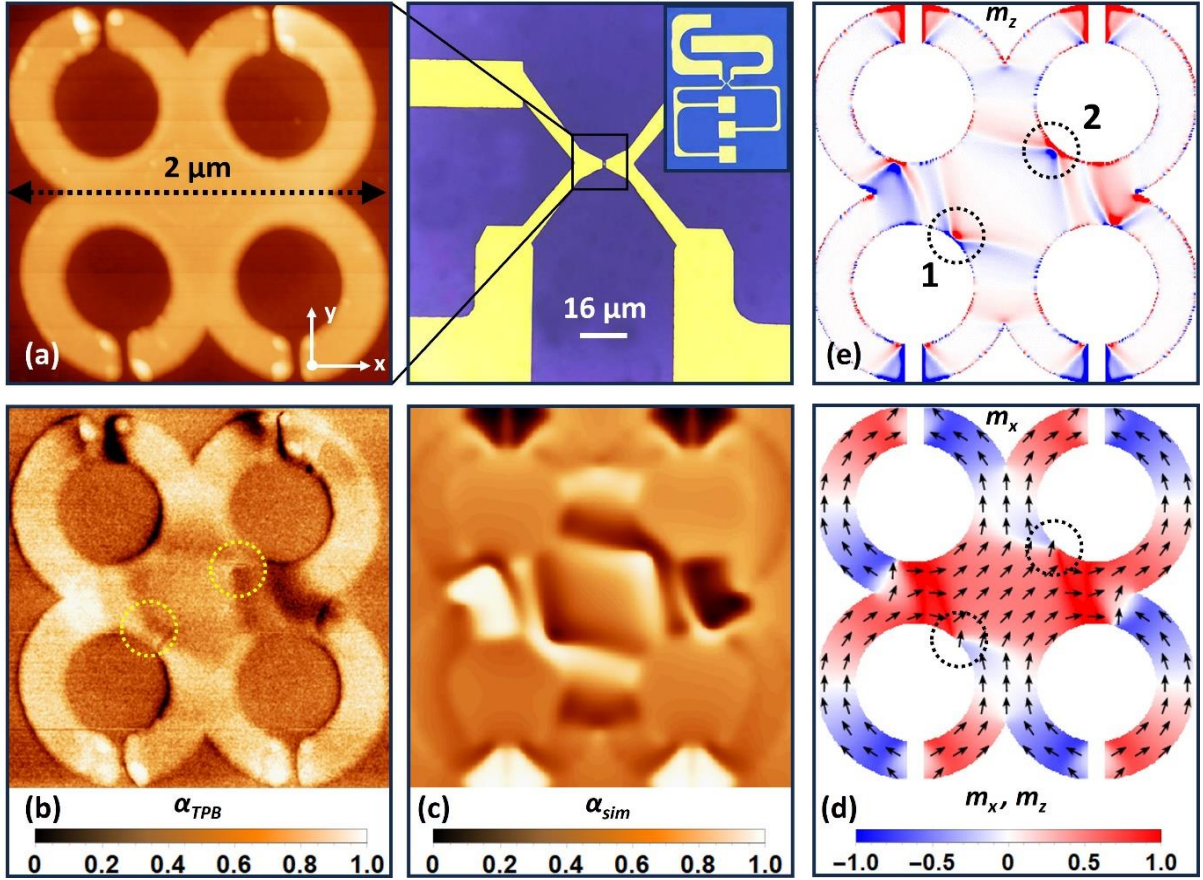

**FIG. 1.** (a) AFM image of the fabricated device in the shape of truncated astroid. (b) MFM image of astroid device after magnetic field conditioning, obtained from taping phase backward (TPB) channel. The gradient of magnetization is determined by the change in the phase (normalized to 0-1 range for better comparison with simulations) of the oscillating tip during magnetic interactions. The yellow circles indicate where AV-core nucleates. (c) Simulated MFM image corresponding to the magnetic remanent state shown in (b). (d, e) Magnetic configuration at remanence obtained after magnetic field conditioning as computed from micromagnetic simulations. The color scale indicates the x-component of magnetization (d), z-component of magnetization (e), the black arrows describe the in-plane component of the magnetization. The black circles highlight AV-core locations 1 and 2.

After verification of the antivortex state at remanence using magnetic imaging, we have made electrical contacts to the bonding pads on two opposite sides of the device along the x-axis (see supplementary Fig. S1-b) using Au wire bonds. We have monitored the device resistance while scanning the magnetic field along the y-axis using a lock-in technique inside the physical property measurement system (PPMS). The device resistances at the remanent state varied between 120 and 150 Ω, and during the field scan a reproducible magnetoresistance response was observed (see Fig. 2). This field dependent resistance change can be attributed to the AMR effect which is a measure of the relative orientation of the magnetization with respect to the electric current. The normalized AMR signal is defined as

$$\frac{\rho - \rho_\perp}{\rho_\parallel - \rho_\perp} = \cos^2(\theta), \tag{1}$$

where $\theta$ is the angle between the magnetization vector and current, $\rho$ is the electrical resistivity, $\rho_\parallel$ is the longitudinal resistivity ($\theta = 0°, 180°$) and $\rho_\perp$ is the perpendicular resistivity ($\theta = 90°, 270°$). When the device is saturated along the y-axis/x-axis with a high magnetic field, low/high resistance state implying current perpendicular/parallel to magnetization is observed.

Figure 2 shows the normalized AMR signals measured at T = 300 K (grey), 200 K (purple), 100 K (orange) and the simulated AMR response at 0 K (blue). In order to facilitate the comparison, all curves have been offset vertically. The inset shows the rectangular central region where the AMR signal acts as a local probe of the relative orientation of the magnetization with respect to +x direction. First, we saturate the magnetization by applying a 500 Oe magnetic field in +y direction. Then, we monitor the device resistance as a function of the magnetic field as we scan the field between +500 Oe and –500 Oe to complete the AMR loop (see supplementary material for AMR measurement details). The same procedure is applied in simulations.

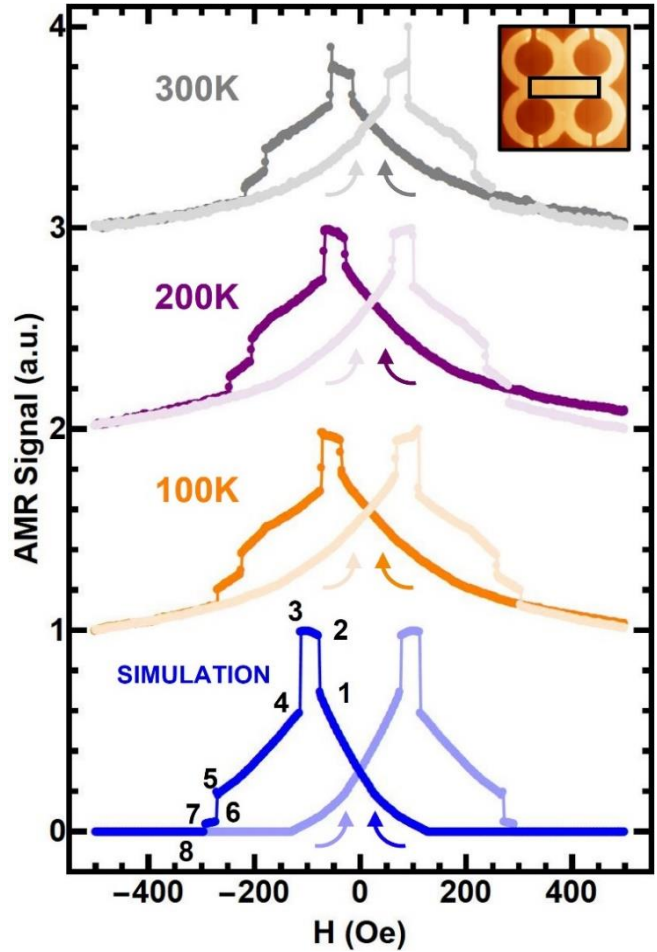

**FIG. 2.** Normalized AMR signal as a function of applied field at T = 300 K (grey), 200 K (purple), 100 K (orange), and simulated AMR response at 0 K (blue). Resistance transitions are labeled on the simulated curve. Inset shows the current flow region as a rectangular box centered on a truncated astroid device. Arrows indicate the scanning direction of the magnetic field.

At 100 K, decreasing the magnetic field from +500 Oe to +150 Oe, AV-cores slowly enter the central region, at –66 Oe (-76 Oe 0 K simulation) a switching to a high resistance state occurs. At –109 Oe (-112 Oe 0 K simulation) switching back to a lower resistance state occurs. At –258 Oe and –301 Oe (-270 Oe and –292 Oe 0 K simulations) there are two other switching events. At higher temperatures the AMR loops look structurally the same only with slightly (~10 Oe) lower switching fields. The micromagnetic simulation snapshots shown in Figure 3 allow us to interpret the observed switching events as 1-2 switching from diagonal AV-core configuration to in-plane uniform magnetization, 3-4 nucleation of AV-cores in the opposite diagonal configuration, 5-6 annihilation of the top-left AV, 7-8 annihilation of the bottom-right AV. In micromagnetic simulations, the applied magnetic field direction is slightly (1°) misaligned from the +y axis to break the symmetry for convergence and similarly experimentally the applied magnetic field is misaligned (~1-2°) due to imperfect sample mounting. This can explain the small difference in switching fields for the transitions 5 to 6 and 7 to 8. Other potential sources of the small differences in switching fields include edge defects, sample imperfections etc.

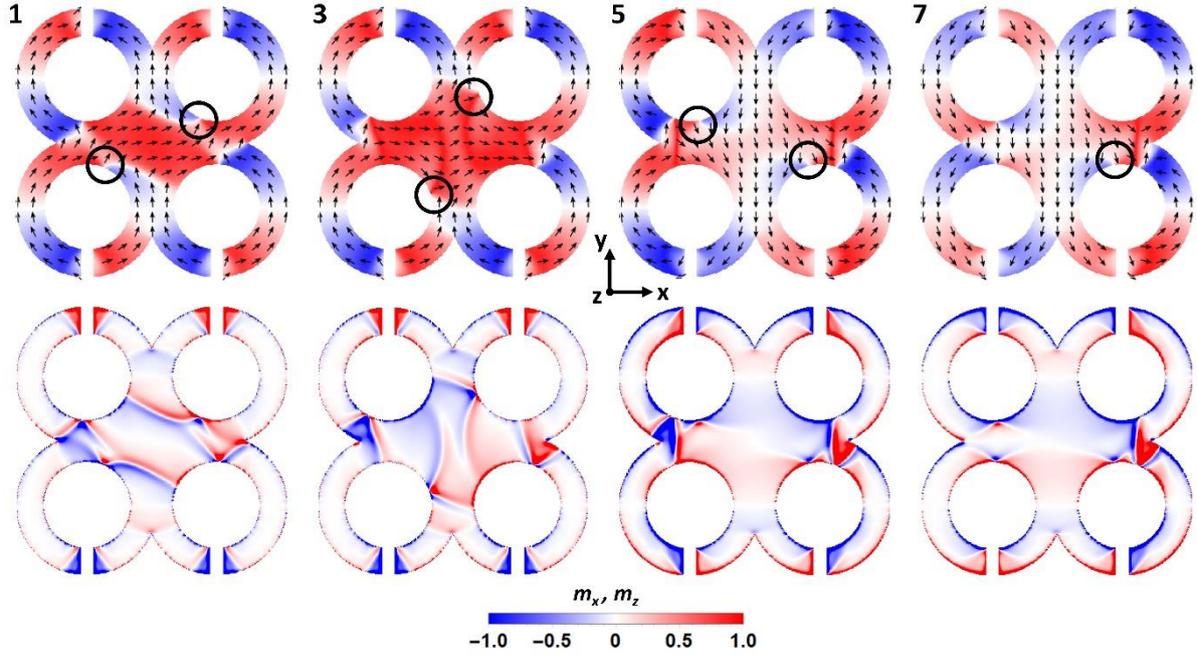

**FIG. 3.** Snapshots of the magnetic configurations corresponding to the states labeled 1, 3, 5, and 7 in Figure 2 (for the states of 2, 4, 6, and 8 see the supplementary Fig. S2). Color scale shows the x-component (top) and z-component (bottom) of the magnetization. The arrows indicate the in-plane components.

To observe zero field AV-core dynamics driven by spin-torque at room temperature, we measure the frequency spectra as a function of DC current using the setup shown in supplementary Fig. S1-c. The DC current flows in +x-direction and the RF component of the device signal is filtered out using a bias tee. The RF signal is amplified using a low noise RF amplifier and detected by spectrum analyzer. Figure 5-a shows the observed resonance peaks for applied current densities in the range of 1 - 1.2 × $10^{10}$ A/m². The resonance peak appears to be blue-shifted from 1.158 to 1.168 GHz with increasing current. The output power is approximately 80 pW. Attempts to increase the current density further resulted in irreversible damage to the device due to excessive heating. Figure 4-b shows J = 1.2 ×$10^{10}$ A/m² resonance peak linewidth determination using a Lorentzian fit yielding FWHM of 23.3 kHz and quality factor ~50000.

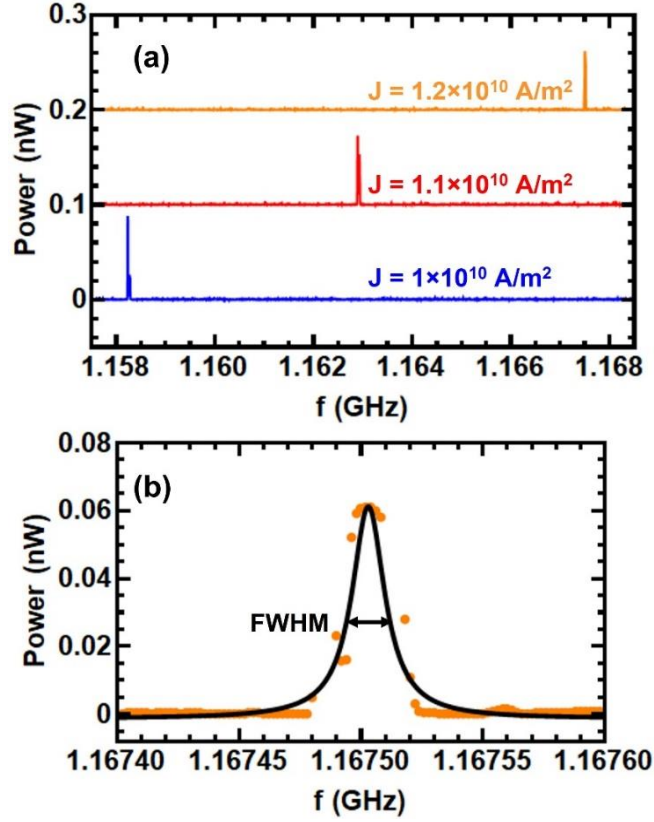

**FIG. 4.** (a) Spin-torque induced RF oscillations spectra (RF Power vs. Frequency) corresponding to current densities J = 1×10$^{10}$ A/m$^2$ (blue), 1.1×10$^{10}$ A/m$^2$ (red), 1.2×10$^{10}$ A/m$^2$ (orange). Spectra are offset for clarity. (b) Close-up view of the spectrum corresponding to J = 1.2×10$^{10}$ A/m$^2$ together with Lorentzian fit (black) for determination of the linewidth (FWHM = 23.3 kHz).

In pursuit of both AV-core stabilization criteria and a clear picture of the spin-torque induced magnetization dynamics, we have performed micromagnetic simulations based on the numerical solution of the Landau-Lifshitz-Gilbert-Slonczewski (LLGS) equation [31-33] using OOMMF 2.0a3 [29] excluding finite temperature effects. The material parameters used in simulations are listed in supplementary material. The Oersted field generated by the DC current was calculated using the formalism described by Ref. 35. Considering the comparable measured resistivities of Py layer and Pt layer (~20-25 μΩ.cm) with a thickness ratio of 10:1, approximately 10 % of the applied current flows through the Pt layer where transverse spin current is generated via the Spin Hall Effect (SHE). The spin-torque that develops between the spin current and the local magnetic moment is included by using an effective spin-polarization of 0.6 in +y direction [35].

Figure 5 shows the micromagnetic simulation results of the spin-torque induced persistent oscillations triggered at current densities above $3\times10^{10}$ A/m$^2$. These simulations reveal that the observed oscillations are due to a non-uniform excitation of one of the AV-cores (bottom left see Fig. 1-d) coupled to a closure domain wall resulting in two domain walls propagating on a curved trajectory as shown by the yellow arrows in Figure 5-a inset (see the supplementary movies). Indeed, the top right AV-core remains stationary until getting disturbed by spin waves emitted from the bottom left corner for current densities $J \geq 12\times10^{10}$ A/m$^2$. The simulations suggest that the threshold current density is $3\times10^{10}$ A/m$^2$ which is a factor of 3 higher than the experimental value $1\times10^{10}$ A/m$^2$ implying that these oscillations are thermally assisted and steady (see Fig. 5-b). For low current densities $3$-$4\times10^{10}$ A/m$^2$, the resonance peaks occur at 1.13-1.18 GHz with relatively low power (see Fig. 5 c-d low J region) and increase with increasing current density in good agreement with the experimental findings (1.16-1.17 GHz, blue-shifted with increasing current). The transition from low power to high power output is related to the amplitude of domain-wall oscillations such that for low J, the oscillations occur within the central region and for high J the oscillation amplitude increases causing a periodic entrance and exit of the domain wall in and out of the central current flow region. Since the high J region is experimentally inaccessible due to excessive heating caused damage, this transition was not detected.

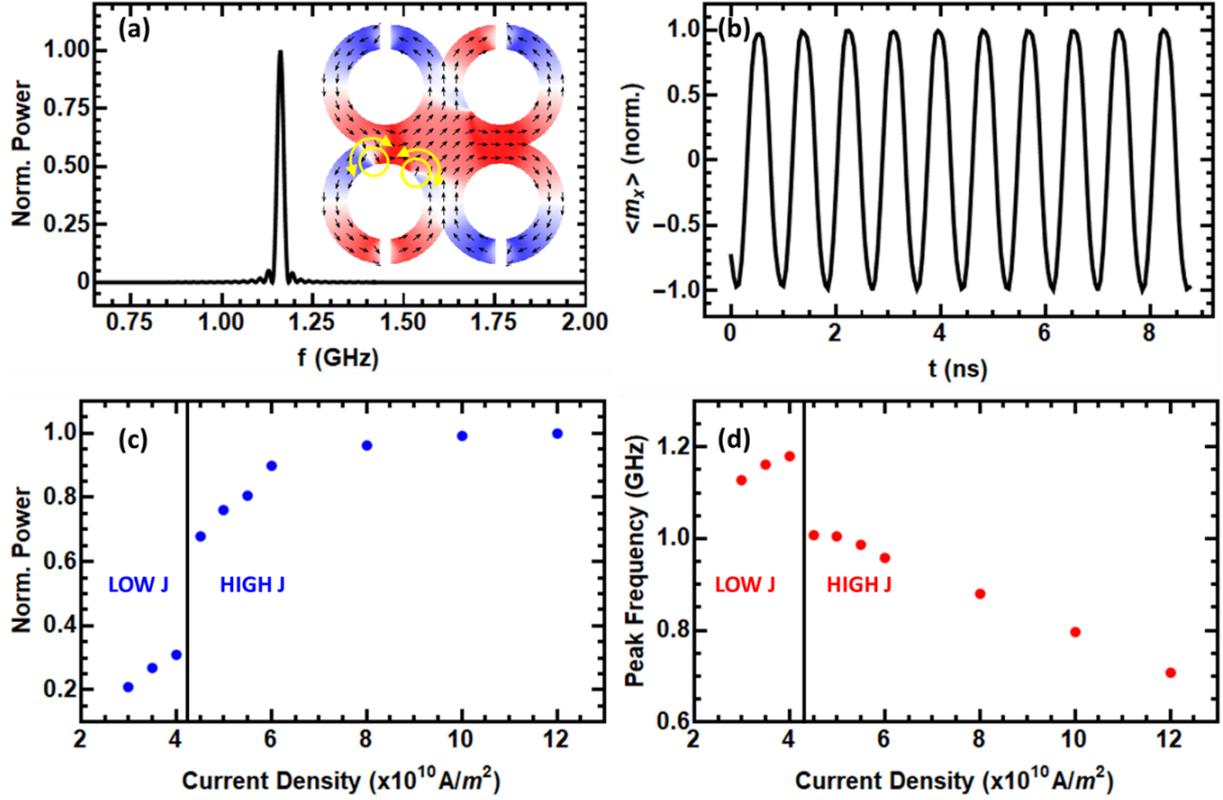

**FIG. 5.** Micromagnetic simulation results for spin-torque induced AV-core oscillations. (a) RF oscillations spectrum (normalized RF power vs. frequency) for $J = 3.5 \times 10^{10}$ A/m$^2$. Inset is the t = 3.5 ns snapshot of micromagnetic configuration after the onset of persistent oscillations. The closure domain wall and the AV-core are highlighted with yellow circles on left and right, respectively. The arrows indicate the domain wall trajectory during oscillations (see the supplementary movies) (b) Time evolution of the normalized x-component of magnetization for $J = 3.5 \times 10^{10}$ A/m$^2$. (c) Normalized power vs. current density. (d) Peak frequency vs. Current density. Low J region corresponds to $J \leq 4 \times 10^{10}$ A/m$^2$.

We have performed detailed analysis of micromagnetic simulations by considering the effect of Oersted field and growth induced uniaxial anisotropy. Comparing measured spectra with the simulated spectra, while the resonance peak locations and current dependent shifts are quite similar in the low J region, the experimental threshold current density is factor of 3 lower. We attribute this reduction to thermally assisted excitations that were not taken into account at zero temperature simulations.

Since the current flows laterally, there are in principle two sources of spin-torque; one originating from spin-polarized lateral current i.e. spin-transfer-torque (STT) [36] and the other due to transverse pure spin current generated at the Pt-Py interface i.e. spin-orbit-torque [37]. By setting the spin-polarization in +y direction in the simulations which would be determined by the positive spin Hall angle in Pt [38], the close agreement obtained between the simulations and the experiment implies that these two torque terms act in a similar way. Besides, another reason behind the experimental lower threshold current could be this additional self-torque not considered in simulations.

Finally, it is worth comparing the RF oscillation characteristics of AV-core obtained in this study with that of vortices obtained in previous studies. The resonance frequency and output power of the AV-core-coupled-to-non-uniform domain wall oscillation mode (1.16-1.17 GHz and tens of pW) is quite similar to the magnetic vortex oscillator core gyration mode (1.1 GHz and same power level) [39]. In contrast, both the linewidth of the AV-core oscillator (20-30 kHz) and the threshold current ($1 \times 10^{10}$ A/m$^2$) is an order of magnitude lower than the vortex oscillator (100-300 kHz, $6 \times 10^{11}$ A/m$^2$) resulting in a quality factor about 4 times higher (QF ~ 50000) with a lower current requirement. The downside of the AV-core oscillator is the necessity to pattern a more complicated device (the astroid geometry) as opposed to a single disk. However, the choice of SOT instead of STT enables much simpler contact geometry in the current in-plane configuration. This advantage of SOT was also recently utilized in vortex oscillators [25].

In conclusion, we were able to observe stable AV-core nucleation in a special geometry, where the current flow path was carefully designed so that spin-torque could effectively drive the persistent oscillations of half-antivortex cores reminiscent of an isolated AV. AMR measurements together with micromagnetic simulations lead to a clear detection of AV-core nucleation/annihilation events also verified by direct magnetic imaging with MFM. By utilizing SHE on a Pt cap layer a transverse spin-current is generated and injected into the neighboring Py layer developing a mutual spin-torque between the local moments and the spin current. Above the threshold current density of $1 \times 10^{10}$ A/m$^2$ this local interaction can effectively trigger GHz frequency oscillations with a quality factor as high as 50000. The source of the signal through micromagnetic simulations was diagnosed as non-uniform excitation of one of the AV-cores coupled to a closure domain wall propagating on a curved trajectory. While the output power of a single device is in the 10s of pW range, high power signal can potentially be obtained using phase-locking in a device array with Magnetic Tunnel Junction readers [15,40,41]. This fundamental study of antivortex core response to spin currents is crucial for the assessment of their potential applications in high frequency spintronic devices and neuromorphic computing applications.

See the supplementary material for detailed descriptions of the sample fabrication, AMR measurement setup, field conditioning procedure, and micromagnetic simulations.

Work at Argonne was supported by the U.S. Department of Energy, Office of Science, Materials Science and Engineering Division, Basic Energy Sciences (BES). Work performed at the Center for Nanoscale Materials, a U.S. Department of Energy Office of Science User Facility, was supported by the U.S. DOE, Office of Basic Energy Sciences, under Contract No. DE-AC02-06CH11357. O. O. and V. K. acknowledge support from Bogazici University Research Fund (17B03D3), TUBITAK 2214/A, and the U.S. Department of State Fulbright Visiting Scholar Program. Work at Bogazici University was supported by Bogazici University Research Fund BAP project number 16B03P5 and TUBITAK project number 113F378. The authors thank Sabanci University SUNUM for nanofabrication support.

## AUTHOR DECLARATIONS

### Conflict of Interest

The authors have no conflicts to disclose.

### Author Contributions

**Ahmet Koral Aykin:** Formal analysis (equal); Investigation (lead); Methodology (equal); Validation (lead); Writing – original draft (lead); Writing – review & editing (equal). **Hasan Piskin:** Formal analysis (equal); Investigation (equal); Methodology (equal); Validation (lead); Writing – original draft (lead); Writing – review & editing (equal). **Bayram Kocaman:** Investigation (supporting); Writing – review & editing (supporting). **Cenk Yanik:** Investigation (equal). **Mario Carpentieri:** Investigation (supporting). **Giovanni Finocchio:** Conceptualization (equal); Formal analysis (equal); Methodology (equal); Validation (equal); Investigation (supporting); Writing – review & editing (supporting). **Ozhan Ozatay:** Conceptualization (lead); Formal analysis (lead); Methodology (lead); Validation (equal); Writing – original draft (lead); Writing – review & editing (lead). **Vedat Karakas** Investigation (supporting); Methodology (equal). **Sevdenur Arpaci:** Investigation (supporting); Methodology (equal). **Aisha Gokce Ozbay:** Investigation (supporting); Methodology (equal). **Federica Celegato:** Methodology (equal). **Paola Tiberto:** Methodology (equal). **Sergi**

**Lendinez:** Formal analysis (equal); Methodology (equal); Investigation (supporting). **Valentine Novosad:** Methodology (equal); Investigation (supporting). **Axel Hoffmann:** Conceptualization (lead); Investigation (supporting); Methodology (equal); Validation (lead).

## DATA AVAILABILITY

The data that support the findings of this study are available from the corresponding author upon reasonable request.

## References

1. C. E. Zaspel, "Magnetic antivortex dynamics in a two nanocontact disk," J. Appl. Phys. 121, 213906 (2017).

2. K. Tanabe and K. Yamada, "Influence of the Dzyaloshinskii–Moriya interaction on the topological Hall effect in crossed nanowires," Appl. Phys. Express 11, 113003 (2018).

3. A.P. Malozemoff and J.C. Slonczewski, "Magnetic Domain Walls in Bubble Materials," Academic Press, New York, 1979.

4. M. Goto, Y. Nozaki and K. Sekiguchi, "Criteria for electric determination of antivortex creation in ferromagnetic thin film," Jpn. J. Appl. Phys. 54, 023001 (2015).

5. M. Pues, M. Martens and G. Meier, "Absorption spectroscopy of isolated magnetic antivortices," J. Appl. Phys. 116, 153903 (2014).

6. S. Gliga, R. Hertel and C. M. Schneider, "Switching a magnetic antivortex core with ultrashort field pulses," J. Appl. Phys. 103, 07B115 (2008).

7. K. Shigeto, T. Okuno, Ko Mibu, T. Shinjo and T. Ono, "Magnetic force microscopy observation of antivortex core with perpendicular magnetization in patterned thin film of permalloy," Appl. Phys. Lett. 80, 4190 (2002).

8. A. Lara, O. V. Dobrovolskiy, J. L. Prieto, M. Huth, F. G. Aliev, "Magnetization reversal assisted by half antivortex states in nanostructured circular cobalt disks," Appl. Phys. Lett. 3, 182402 (2014).

9. H. Wang and C. E. Campbell, "Spin dynamics of a magnetic antivortex: Micromagnetic simulations," Phys. Rev. B 76, 220407(R) (2007).

10. A. Drews, B. Krüger, G. Meier, S. Bohlens, L. Bocklage, T. Matsuyama and M. Bolte, "Current- and field-driven magnetic antivortices for nonvolatile data storage," Appl. Phys. Lett. 94, 062504 (2009).

11. T. Kamionka, et al., "Magnetic antivortex-core reversal by circular-rotational spin currents," Phys. Rev. Lett. 105, 137204 (2010).

12. M. Pues, M. Martens, T. Kamionka and G. Meier, "Reliable nucleation of isolated magnetic antivortices," Appl. Phys. Lett. 100, 162404 (2012).

13. A. Haldar and K. S. Buchanan, "Magnetic antivortex formation in pound-key-like nanostructures," Appl. Phys. Lett. 102, 112401 (2013).

14. M. Asmat-Uceda, L. Li, A. Haldar, B. Shaw and K. S. Buchanan, "Geometry and field dependence of the formation of magnetic antivortices in pound-key-like structures," J. Appl. Phys. 117, 173902 (2015).

15. A. Ruotolo, V. Cros, B. Georges, A. Dussaux, J. Grollier, C. Deranlot, R. Guillemet, K. Bouzehouane, S. Fusil and A. Fert, "Phase-locking of magnetic vortices mediated by antivortices," Nature Nanotech 4, 528 (2009).

16. G. Finocchio, O. Ozatay, L. Torres, R. A. Buhrman, D. C. Ralph and B. Azzerboni, "Spin-torque-induced rotational dynamics of a magnetic vortex dipole," Phys. Rev. B 78, 174408 (2008).

17. C. Ragusa, M. Carpentieri, F. Celegato, P. Tiberto, E. Enrico, L. Boarino, G. Finocchio, "Magnonics crystal composed by magnetic antivortices confined in antidots," IEEE Trans. Mag. 47, 2498 (2011).

18. A. Drews, B. Krüger, M. Bolte and G. Meier, "Current- and field-driven magnetic antivortices," Phys. Rev. B 77, 094413 (2008).

19. T.-J. Meng, J.-B. Laloë, S. N. Holmes, A. Husmann and G. A. Jones, "In-plane magnetoresistance and magnetization reversal of cobalt antidot arrays," J. Appl. Phys. 106, 033901 (2009).

20. M. Goto and Y. Nozaki, "Current-driven antivortex core resonance measured by the rectifying effect," AIP Advances 6, 025313 (2016).

21. G. A. Riley, H. J. Jason Liu, M. A. Asmat-Uceda, A. Haldar and K. S. Buchanan, "Observation of the dynamic modes of a magnetic antivortex using Brillouin light scattering," Phys. Rev. B 92, 064423 (2015).

22. C.E. Zaspel, "Antivortex gyrotropic motion in a nanocontact," J. Magn. Magn. Mater. 323, 499 (2011).

23. X. J. Xing, Y. P. Yu, S. X. Wu, L. M. Xu and S. W. Lib, "Bloch-point-mediated magnetic antivortex core reversal triggered by sudden excitation of a suprathreshold spin-polarized current," Appl. Phys. Lett. 93, 202507 (2008).

24. X.-J. Xing and S.-W. Li, "Spin-transfer torque driven magnetic antivortex dynamics by sudden excitation of a spin-polarized dc," J. Appl. Phys. 105, 093902 (2009).

25. S. Lendínez, T. Polakovic, J. Ding, M. B. Jungfleisch, J. Pearson, A. Hoffmann and V. Novosad, "Temperature-dependent anisotropic magnetoresistance and spin-torque-driven vortex dynamics in a single microdisk," J. Appl. Phys. 127, 243904 (2020).

26. M. N. Dubovik, V. V. Zverev and B. N. Filippov, "Nonlinear dynamics of domain walls with cross-ties," J. Exp. Theor. Phys. 123, 108 (2016).

27. A. Brataas, A. D. Kent and H. Ohno, "Current-induced torques in magnetic materials," Nat. Mater. 11 ,372-381 (2012).

28. D. Pinna, G. Bourianoff, and K. Everschor-Sitte, "Reservoir Computing with Random Skyrmion Textures," Phys. Rev. Applied 14, 054020 (2020).

29. M. Donahue, "OOMMF User's Guide, Version 1.0," - 6376, National Institute of Standards and Technology, Gaithersburg, MD (1999).

30. J. Brandão, S. Azzawi, A. T. Hindmarch and D. Atkinson, "Understanding the role of damping and Dzyaloshiskii-Moria interaction on dynamic domain wall behaviour in platinum-ferromagnet nanowires," Sci Rep 7, 4569 (2017).

31. J. C. Slonczewski, "Current-driven excitation of magnetic multilayers," J. Magn. Magn. Mater. 159, L1-L7 (1996).

32. J. Xiao and A. Zangwill, "Boltzmann test of Slonczewski's theory of spin-transfer torque," Phys. Rev. B 70, 172405 (2024).

33. A. Giordano, V. Puliafito, L. Torres, M. Carpentieri, B. Azzerboni and G. Finocchio, "Micromagnetic study of spin-transfer-driven vortex dipole and vortex quadrupole dynamics," IEEE Trans. Mag. 50, 4300404 (2014).

34. M. Hayashi, J. Kim, M. Yamanouchi and H. Ohno, "Quantitative characterization of the spin-orbit torque using harmonic Hall voltage measurements," Phys. Rev. B 89, 144425 (2014).

35. O. Ozatay, N. C. Emley, P. M. Braganca, A. G. F. Garcia, G. D. Fuchs, I. N. Krivorotov, R. A. Buhrman, and D. C. Ralph, "Spin transfer by nonuniform current injection into a nanomagnet," Appl. Phys. Lett. 88, 202502 (2006).

36. S. Zhang and Z. Li, "Roles of nonequilibrium conduction electrons on the magnetization dynamics of ferromagnets," Phys. Rev. Lett. 93, 127204 (2004).

37. A. Manchon, J. Železný, I. M. Miron, T. Jungwirth, J. Sinova, A. Thiaville, K. Garello, P. Gambardella, "Current-induced spin-orbit torques in ferromagnetic and antiferromagnetic systems," Rev. Mod. Phys. 91, 035004 (2019).

38. L. Liu, T. Moriyama, D. C. Ralph and R. A. Buhrman, "Spin-torque ferromagnetic resonance induced by the spin hall effect," Phys. Rev. Lett. 106, 036601 (2011).

39. V. S. Pribiag, I. N. Krivorotov, G. D. Fuchs, P. M. Braganca, O. Ozatay, J. C. Sankey, D. C. Ralph and R. A. Buhrman, "Magnetic vortex oscillator driven by d.c. spin-polarized current," Nature Phys 3, 498 (2007).

40. S. Wittrock, P. Talatchian, M. Romera, S. Menshawy, M. J. Garcia, M-C. Cyrille, R. Ferreira, R. Lebrun, P. Bortolotti, U. Ebels, J. Grollier and V. Cros, "Beyond the gyrotropic motion: Dynamic C-state in vortex spin torque oscillators," Appl. Phys. Lett. 118, 012404 (2021).

41. S. Sani, J. Persson, S. M. Mohseni, Y. Pogoryelov, P. K. Muduli, A. Eklund, G. Malm, M. Käll, A. Dmitriev and J. Åkerman, "Mutually synchronized bottom-up multi-nanocontact spin–torque oscillators," Nat Commun 4, 2731 (2013).